\documentclass[aps,prb,twocolumn,no-footinbib,showpacs,floatfix,amsmath,amssymb]{revtex4-1}

\usepackage{graphicx}
\usepackage{dcolumn}
\usepackage{bm}
\usepackage[normalem]{ulem}
\usepackage{subfigure} 
\usepackage{soul,xcolor}
\usepackage{siunitx}
\usepackage{textcomp}
\usepackage{float} 

\setstcolor{red}
\RequirePackage{color}

\begin{document}

\title{Time-diffraction and Zitterbewegung of two-dimensional massless Dirac excitations}
\author{Elmer~Cruz}
\author{Ramon~Carrillo-Bastos}
\author{Jorge~Villavicencio}
\email{villavics@uabc.edu.mx}
\affiliation{Facultad  de  Ciencias,  Universidad  Aut\'onoma  de  Baja  California, 22800  Ensenada,  Baja  California,  M\'exico.}
\date{\today}

\begin{abstract}

We explore the dynamics of two-dimensional massless Dirac-fermions  within a quantum shutter approach, which involves the time-evolution of an initial cut-off plane wave.
We show that the probability density is governed by an interplay between {\it diffraction in time} and  {\it Zitterbewegung} phenomena, typical of relativistic quantum shutter systems with nonzero mass.  
The {\it time-diffraction}  appears as an oscillatory pattern in the probability density, similar to the effect predicted by Moshinsky in 1952 [Phys. Rev. \textbf{88}, 625] for Schr\"odinger free matter-waves.
The {\it Zitterbewegung}  manifests itself as high-frequency oscillations embedded in the time-diffraction profile. 
We found that these two transient effects are induced by the transverse momentum component of the incident wave, $k_y$, that acts as an effective mass of the system.
Furthermore, this effective  mass can be manipulated by tuning the incidence angle of the initial quantum state, which allows to control the frequencies of the transients. In particular, we demonstrate that near a normal incidence condition, the  {\it Zitterbewegung} appears as a 
series of {\it quantum beats} in the probability density, with a beating frequency $2k_yv_F$, where $v_F$ is the Fermi velocity.

\end{abstract}
\maketitle


Since the experimental discovery of graphene\cite{Novoselov666}, the study of two-dimensional (2D) Dirac fermions in condensed matter physics has become of great interest  both at the fundamental and applied levels \cite{KATSNELSON200720}. Graphene exhibits a dispersion relation  with two bands touching at two singular points at the corners of the Brillouin zone.
Near these Dirac points the spectrum is linear in momentum, and the electronic excitations are described by a Dirac-like equation for massless particles\cite{RevModPhys.81.109,Novoselof05}. Nevertheless, graphene is not the only condensed matter system that exhibits 
an effective Dirac
Hamiltonian\cite{Review-DiracMatter}, other examples include silicene\cite{rise-of-silicene}, germanene\cite{germanene}, \textit{Pmmn} borophene\cite{pmmn-boron}, cold atoms systems\cite{cold-atoms}, and topological insulators\cite{topological-insulators}. These materials constitute the so call Dirac matter, and due to its relativistic Hamiltonian they have allowed probing several phenomena  originally predicted by quantum electrodynamics,  
such as  Klein tunneling\cite{klein1,Klein2} and the {\it Zitterbewegung} (ZBW) effect \cite{Katsnelson2006, frolova2008wave,PhysRevB.76.195439,Krueckl_2009}, which have now become available to experimentalists\cite{Klein3,Gerri}. 
While the Klein tunneling occurs at equilibrium, the ZBW being an oscillation is an out of equilibrium effect,
and in fact it appears as a transient phenomenon in two-dimensional Dirac systems\cite{PhysRevB.76.195439,frolova2008wave}. 
In general, transient effects \cite{k94, dcgcjm, mohsen2013quantum} arise from the properties of quantum waves involving either relativistic or non-relativistic equations subject to sudden changes as initial condition. 
The most representative transient effect is the {\it diffraction in time} (DIT) of free matter-waves, predicted by Moshinsky \cite{mm52} (\citeyear{mm52}) using a non-relativistic plane wave quantum-shutter model. 
A quantum-shutter setup involves a cut-off plane wave representing a matter-wave beam initially confined in a region of space ($x<0$), stopped by an absorbing shutter at $x=0$. After opening  the shutter at $t=0$, the free Schr\"odinger  probability density exhibits a distinctive DIT \cite{mm52, mm76} oscillatory pattern, similar to the intensity profile of a light beam diffracted by a straight edge.
The DIT effect was later verified by multiple condensed matter experiments involving ultracold atoms \cite{dalibard}, cold-neutrons \cite{Hils98}, and atomic Bose-Einstein condensates \cite{colombe05}.
Also, the understanding and control of the features of DIT are of relevance, for example, in the field of atom lasers\cite{Hagley99,tripp00,12290615}, which operate by extracting matter-waves from Bose-Einstein condensates.
After Moshinsky's seminal paper, several theoretical works have addressed the study of transients using various initial quantum states 
as well as generalizations of the shutter model to explore time-dependent phenomena \cite{dcgcjm}.
However, most of these works were developed in the non-relativistic context. The first studies regarding relativistic transients involved solutions to the Klein-Gordon \cite{mm52} and Dirac \cite{moshrmf52} equations for a free massive three-dimensional particle moving in one-dimension within a plane wave quantum-shutter model.
Some later works explored issues such as non-locality, forerunners, and time scales, using different types of relativistic initial conditions \cite{bt98,mbmuga00,dmrgv03,deutchap93,ggcarjvpra99,jpajv00,kalber01,godoy16}. 
A nice feature of these  approaches is that they can describe electronic dynamics based on exact analytical solutions,  which may provide a useful tool for exploring time-dependent features of 2D Dirac-matter. Moreover, although the transport effects in these 2D systems have been studied\cite{PhysRevB.76.195439,frolova2008wave} and the importance of transient phenomena has been stressed, until now (and up to our knowledge) the DIT for two-dimensional massless Dirac particles have not been explored.

In this work we present a new approach based on a quantum shutter model to study transient effects in 2D massless Dirac matter. 
By considering the Hamiltonian for graphene as typical example, we derive an exact analytical solution to explore the systems dynamics, that can be generalize to other 2D Dirac massless materials. 
We find that the probability density of massless Dirac-fermions exhibit two transient effects previously observed (DIT) or overlooked (ZBW) in quantum shutter approaches involving relativistic systems with non-zero mass in vacuum. We argue that these effects are a consequence of a  non-vanishing transverse momenta of the initial quantum state that acts as the mass of the system.

\begin{figure}[H]
 \begin{center}
  {\includegraphics[angle=0,width=3.4in]{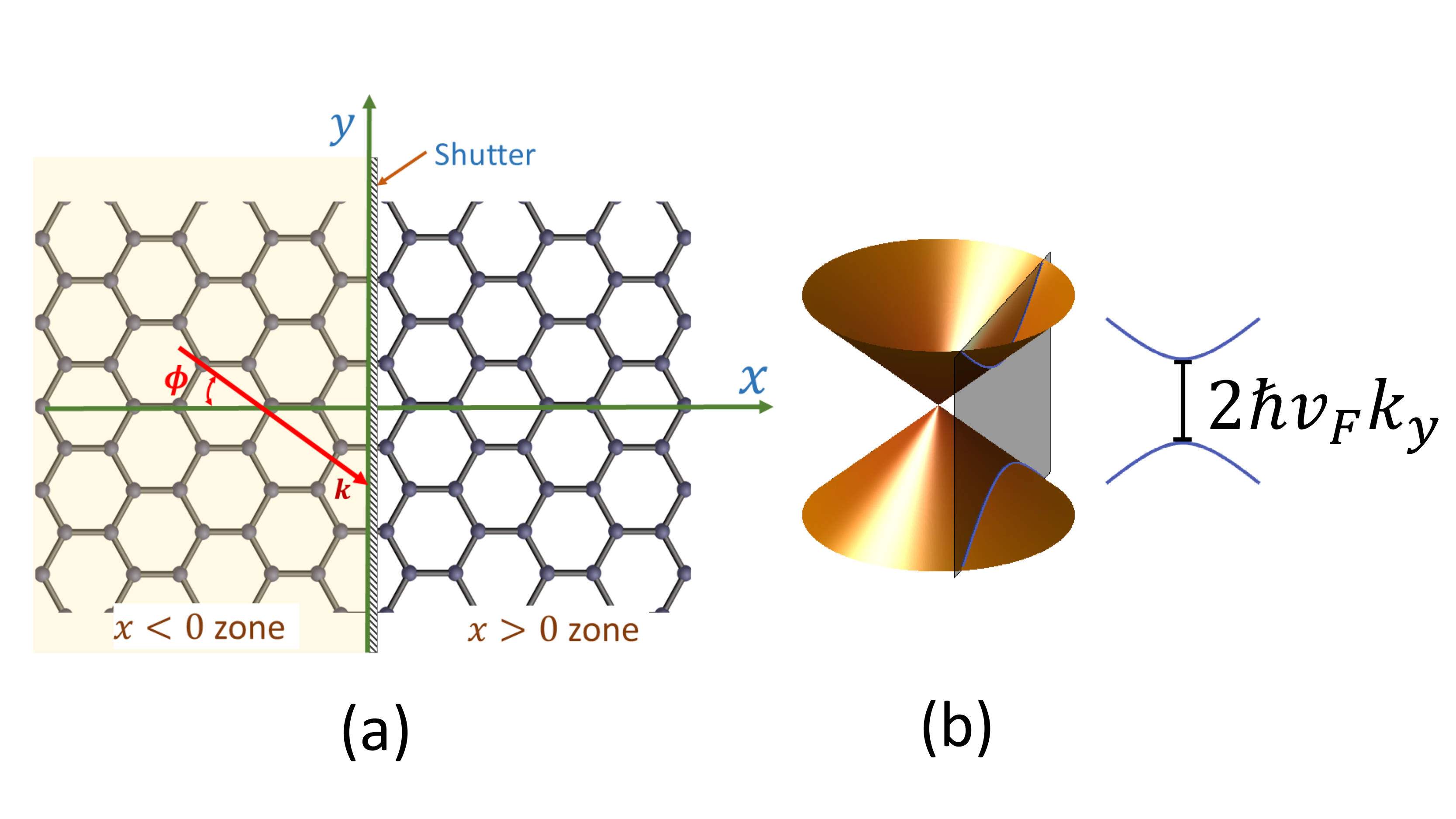}}
   \end{center}
    \caption {(a) Quantum shutter model for electrons on the surface of a
        graphene monolayer. At $t=0$ an initial state with momentum $\bm{k}$, and incidence angle $\phi$, is confined in the region $x\le 0$ by a perfect absorbent shutter located along the $y$-axis. (b) Dirac cone where $k_y$ induces an energy gap that acts as an effective mass of the system.}
          \label{fig:shutcono}
           \end{figure}

\textit{Cut-off plane wave model. } 
As a prototypical example for two-dimensional Dirac matter, we describe the dynamics of low-energy electron excitations in monolayer graphene using a Dirac-like equation,
\begin{equation}\label{Dirac}
i\hbar\,\partial_t\Psi_{\eta}(\boldsymbol{r},t)  = v_{F}\,\boldsymbol{\sigma}_{\eta}\cdot \hat{\boldsymbol{p}}\,\Psi_{\eta}(\boldsymbol{r},t),
\end{equation}
where $v_{F}$ is the Fermi velocity, $\hat{\boldsymbol{p}}=-i\hbar\nabla$ is the momentum operator of the charge carriers, and   $\boldsymbol{\sigma}_{\eta}=\left(\eta\sigma_{x},\sigma_{y} \right)$ is the vector of Pauli matrices. Following Moshinsky's\cite{mm52} approach, we investigate the transient behavior by taking a cut-off two-dimensional plane wave as initial condition, given by, 
\begin{equation}
\Psi_{\eta}(\boldsymbol{r},0)=
e^{i\bm{k}\cdot\bm{r}}\left[\begin{array}{c}
 1 \\
\lambda\,\eta \,e^{i\eta\phi} 
\end{array}\right]\Theta(-x),
\label{cond_ini}
\end{equation}
where $\Theta(-x)$ is a Heaviside function that restricts the plane wave to the semi-infinite plane $(-\infty < x\le 0)$, modeling a perfect absorbing shutter (located at $x=0$), that opens at $t=0$.
The wave vector is $\boldsymbol{k}=(k_x,k_y)$, and the incidence angle $\phi=\tan^{-1}(k_y/k_x)$.
The  Dirac points are labeled by $\eta=\pm1$ (Valley $K$ and $K'$), and $\lambda=\pm1$ is for particles and holes, respectively. See Fig.~\ref{fig:shutcono}.
To solve the latter system we Laplace-transform Eq. (\ref{Dirac}) with the initial condition given by Eq. (\ref{cond_ini}),  and obtain the following time-independent equation,
\begin{equation}\label{Eq:laplaced}
v_{F}\bm{\sigma}_{\eta}\cdot\nabla \widetilde{\psi}_{\eta}(\bm{r},s) +s\widetilde{\psi}_{\eta}(\bm{r},s)=\Psi_{\eta}(\bm{r},0).
\end{equation}
where $\widetilde{\psi}(\boldsymbol{r},s)$  is the  Laplace transformed spinor. 
Here $k_y$ is a good quantum number and thus the solution for the $y$ coordinate will remain a plane wave.
The solutions for the $x$ coordinate must be finite in the limit $|x|\rightarrow \infty$, and continuous at the origin ($x=0$). 
This allows solving the quantum shutter problem in $(x;s)$ space for $x \ge 0$, which can be transformed back to the time-domain by explicitly computing the Bromwich inversion integrals following the procedure described in Refs.  \onlinecite{ggcarjvpra99} and \onlinecite{jpajv00}.
Hence, the time-dependent solution can be written as
\begin{equation}\label{eq8b}
\boldsymbol{\Psi}=
\left[\begin{array}{c}
\Phi^{\uparrow}(x,t)\\
\Phi^{\downarrow}(x,t)
\end{array}\right]\,e^{i k_y y}\,\Theta(v_Ft-x),
\end{equation}
with the spinors $\Phi^{\uparrow\downarrow }(x,t)$ given by,
\begin{eqnarray}
\Phi^{\uparrow}(x,t)&=&(\alpha^+\phi_{+}+\alpha^{-}\phi_{-})-\lambda \,e^{i\eta \phi}J_0(\mu)/2; \label{phiarriba}\\
\Phi^{\downarrow}(x,t)&=&(\beta^+\phi_++\beta^-\phi_{–})-\eta \,J_0(\mu)/2.
\label{phiabajo}
\end{eqnarray}
In Eqs.~(\ref{phiarriba}) and (\ref{phiabajo}), 
$\phi_{\pm}$ resemble the free-type solutions of the Klein-Gordon shutter problem \cite{mm52},   
\begin{equation}
\phi_{\pm}= e^{i[\pm k_x x- \lambda k v_F t]}+\frac{1}{2}%
J_0(\mu ) 
-\sum\limits_{n=0}^\infty (\xi /iz_{\pm })^nJ_n(\mu ),
\label{simplifbis2}
\end{equation}
where $z_{\pm}=(\lambda k\pm k_x)/k_y$,  $\xi=[(v_F t+x)/(v_F t-x)]^{1/2}$, and  $J_{n}(\mu)$ are the Bessel functions of integer order $n$ with argument  $\mu=k_y[v_F^2 t^2 -x^2]^{1/2}$. 
We have also defined the coefficients,
\begin{eqnarray}
\alpha^\pm&=&\frac 1 2\left(-1 \pm i\eta\frac{ k_y}{ k_x} \mp\frac{k}{k_x}e^{i\eta\phi}\right); \\
\beta^\pm&=&-\frac{\lambda}{2}\left[ \pm\eta\frac{ k}{k_x} +\left(\eta \pm i\frac{k_y}{ k_x}\right)e^{i\eta\phi}\right]. 
\end{eqnarray}
Equation (\ref{simplifbis2}) may also be written in the alternative form %
\begin{equation}
\phi_{\pm}= \frac{1}{2}J_0(\mu )
+\sum\limits_{n=1}^\infty (-1)^n (iz_\pm/\xi)^n J_n(\mu), 
\label{pre}
\end{equation}
useful for describing the probability density in the vicinity of   $t=t_F=x/v_F$, as well as the long-time behavior of the relativistic quantum wave. 
We emphasize that in Ref. \onlinecite{mm52} the free Klein-Gordon solution features a DIT effect due to the electron mass, $m_0$. By comparing  Eq.~(\ref{simplifbis2}) with that of Klein-Gordon's model, one may verify that the transverse momentum $k_y$ is equivalent to $\mu_0=m_0 c/\hbar$, which suggests that $k_y$ serves as the mass of the system. 

Our result Eq.~(\ref{eq8b}) can be applied to other two-dimensional Dirac-matter systems, such as 8-$P mmn$ borophene \cite{lopezbonilla16,peng16,Zabolotskiy16,verma17}, 
which exhibits a generalized Dirac dispersion with tilted  cones.
The dynamics in this type of 2D crystalline  structures can be described 
by a massless Dirac Hamiltonian, $\hat{H}=(v_t\sigma_0 \hat{p}_y+v_x\,\sigma_x\, \hat{p}_x+v_y\,\sigma_y \,\hat{p}_y)$,  where $\sigma_i$ are the Pauli matrices, and $\hat{p}_j$ the momentum operator. 
In the continuum model of 8-$P mmn$ borophene, the velocities \cite{Zabolotskiy16} are $v_x = 0.86 v_F$, $v_y = 0.69 v_F$, and $v_t = 0.32 v_F$.
Therefore, we can rewrite the corresponding  Dirac equation $i\hbar \partial_t \psi_b(\boldsymbol{r},t)= \hat{H}\psi_b(\boldsymbol{r},t)$  by using $\psi_{b}(\boldsymbol{r},t)=\Psi_{\eta'}(\boldsymbol{r},t)e^{-iv_tk_yt}$ since $k_y$ is a good quantum number, leading us to
\begin{equation}
i\hbar\,\partial_t\Psi_{\eta'}(\boldsymbol{r},t)  = v_{y}\,\boldsymbol{\sigma}_{\eta'}\cdot \hat{\boldsymbol{p}}\,\Psi_{\eta'}(\boldsymbol{r},t),
\label{borophene_1}
\end{equation}
where $\boldsymbol{\sigma}_{\eta'}=\left(\eta'\sigma_{x},\sigma_{y} \right)$, and $\eta'=v_x/v_y$. 
The equation that describes the dynamics in borophene [Eq.~(\ref{borophene_1})] is very similar to our Eq.~(\ref{Dirac}) for a graphene monolayer, and hence our analytical procedure that led to Eq.~(\ref{eq8b}) can be readily applied. 
We also argue that an alternative procedure for extending our results to other two-dimensional Dirac-matter systems, is to simply transform the Hamiltonian in 
$
-i\hbar \partial_t \Psi'_{\eta}= H'\Psi'_{\eta}
$
via a spin rotation.
That is, by taking $\Psi'_{\eta}=\bm{R}\,\Psi_{\eta}$ with 
$H'=\bm{R}\, \bm{\sigma}_{\eta}\cdot \bm{p}\,\bm{R}^{\dag}$, 
where  $\bm{R}=[\bm{I}\cos(\theta/2)-i\bm{\sigma}\cdot\bm{n}\sin(\theta/2)]$ is the spin-rotation operator.
As usual, $\bm{I}$ is the $2\times 2$ identity matrix, $\bm{\sigma}=(\sigma_x,\sigma_y,\sigma_z)$, and $\bm{n}=(n_x,n_y,n_z)$ is a unitary vector which defines the rotation axis, where $\theta$ is the rotation angle.
Therefore, the transformed Hamiltonian $H'$ can be expressed as
\begin{multline*}
   H'=v_{F}\cos^2(\phi/2)\left(\bm{p}_{\eta}\cdot\bm{\sigma}\right) -v_{F} \sin(\phi)\left[\bm{p}_{\eta}\times\bm{n} \right]\cdot\bm{\sigma}\\ +v_{F}\sin^2(\phi/2)\left(\bm{\sigma}\cdot\bm{n}\right)\left[\bm{p}_{\eta}\cdot\bm{n}+i\left(\bm{p}_{\eta}\times\bm{n}\right)\cdot\bm{\sigma}\right],
\end{multline*}
where $\bm{p}_{\eta}=(\eta p_{x}, p_y)$, 
and by choosing $\phi=0$, the original Hamiltonian $H$ is recovered. 
\begin{figure}[H]
 \begin{center}
  {\includegraphics[angle=0,width=3.5in]{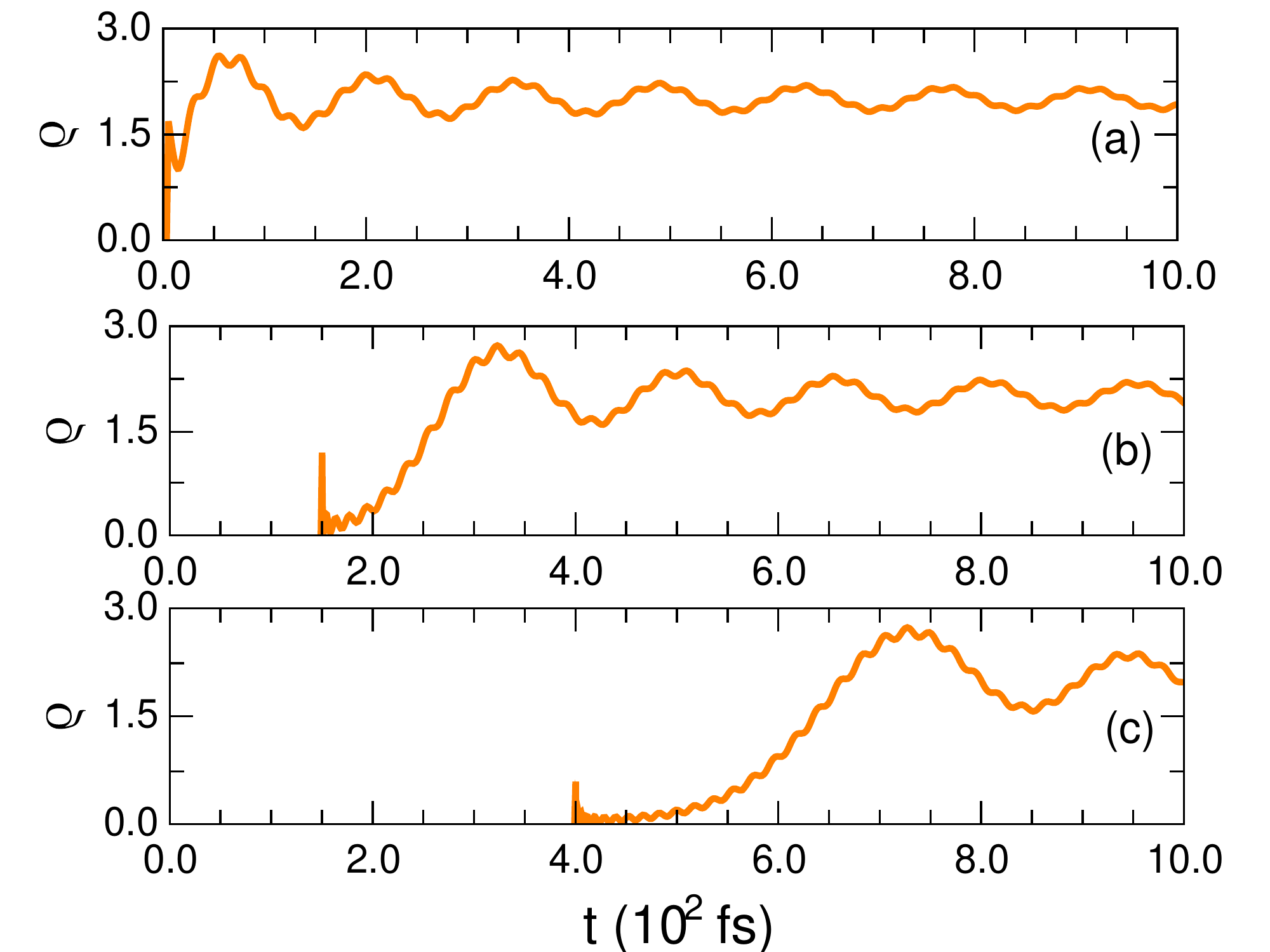}}
    \end{center}
     \caption{Transient behavior of $\rho$ as  function of time $t$, with
      $\phi=\pi/4$, for three fixed values of position (a) $x=5.0$ nm, (b) $x=150.0$ nm, and (c) $x=400.0$ nm. The probability density exhibits large oscillations similar to the DIT phenomenon typical of matter-waves. Here and in all of our calculations we set $\lambda=\eta=1$ due to the symmetry properties of $\rho$, and consider an energy range $E\in[0.05,0.2]$ eV, which is lower than the bandwidth energy ($\sim 3$ eV) (linear momentum approximation), where we choose $E=0.1$ eV, and $v_F=1.0$ nm/fs.}
            \label{fig1}
             \end{figure}
\textit{Transient dynamics in graphene.} 
Given the exact  wavefunction in Eq.~(\ref{eq8b}), we study the dynamics of quantum waves in graphene from the transient to the stationary regime. In particular, we explore the probability density,
$\rho=\boldsymbol{\Psi}^{\dag}\boldsymbol{\Psi}$,
of an electron in the conduction band, localized at the $K$ or $K'$ valley (Dirac point). 
Since $\rho$ is independent of the $y$ coordinate, we shall study $\rho$ as function of position $x$, and time $t$, for different values of the incidence angle, $\phi$.
In Figs.~\ref{fig1}(a)-(c)  we show the time-dependence of $\rho$  for
different fixed values of position, $x$.
We observe that $\rho$, from $t>t_F$ onward, 
grows towards a maximum value from which it oscillates until it reaches the stationary value.
The large period oscillations of $\rho$ shown in Fig.~\ref{fig1} are similar to the DIT phenomenon predicted in Ref. \onlinecite{mm52,mm76} for non-relativistic free matter-waves, which resembles an  intensity pattern of the Fresnel diffraction of light by a semi-infinite plane. 
The connection with optical phenomena is further emphasized since the solutions $\phi_{\pm}$ can also be expressed in terms of Lommel functions of two variables, originally introduced in the context of optical diffraction problems \cite{Watson96}. 
We also notice that the DIT-type profile observed in Fig.~\ref{fig1} exhibits embedded high-frequency oscillations. 
To explain the frequency content of these oscillations, we shall derive an approximate expression for $\rho$. 
We proceed by first noting that for $\phi=\pi/4$, 
the contributions of  $\phi_{-}$ in Eqs.~(\ref{phiarriba}) and (\ref{phiabajo}) cancel out exactly ($\alpha^-=\beta^{-}=0$), allowing a simplification of Eq.~(\ref{eq8b}), which now only depends on $\phi_{+}$.  
By approximating $\phi_{+}$ using the asymptotic expansion for the Bessel functions with large values of the argument,  $J_{n}(z)\simeq (2/\pi z)^{1/2}cos(z-n\pi /2-\pi/4)$, we obtain the probability density $\rho_a=\rho_{dit}+ \rho_{zbw}$, with
\begin{eqnarray}
\rho_{dit}&\simeq&  2-\sqrt{2}\,\gamma\, t^{-1/2} \cos(k_x x-\Omega_{D} t-\pi/4);  \label{rho0_DIT}\\
\rho_{zbw}&\simeq& \left[\left(\sqrt{2}-1\right)\big/\left(\sqrt{2}+1\right)\right] z_+^{-1}\gamma\, t^{-1/2} \sin(k_xx-\omega t) \nonumber \\
&\times& \cos\left[\left(\Omega_Z\,t + \pi/2\right)/2\right],  \label{rho1_ZB}
\end{eqnarray}
where $\omega=k v_F$ is the frequency associated to the initial quantum state, and $\gamma=(8/\pi k_y v_F)^{1/2}$.
In Fig.~\ref{fig1_bis}(a) we show that $\rho_a$ provides a good approximation to the oscillating pattern discussed in Fig.~\ref{fig1}(a). 
Also, from Eqs.~(\ref{rho0_DIT}) and (\ref{rho1_ZB}) the dynamics of $\rho_a$ is governed by two transient contributions, $\rho_{dit}$, and $\rho_{zbw}$. 
In Fig.~\ref{fig1_bis}(b) we show that $\rho_{dit}$ [Eq.~(\ref{rho0_DIT})] describes the DIT pattern, which is characterized by a frequency $\Omega_{D}=(k-k_y)v_F$ in the range of tens of terahertz, 
with a corresponding period $T_D=2\pi\Omega_D^{-1}$ of the order of  femtoseconds.
In Fig.~\ref{fig1_bis}(c) we observe that the high-frequency secondary oscillations embedded in $\rho$, are a manifestation of the ZBW effect,  described by the contribution $\rho_{zbw}$ [Eq.~(\ref{rho1_ZB})]. These oscillations are characterized by a ZBW frequency $\Omega_{Z}=2k_y v_F$, in the range of hundreds of terahertz, with a period $T_Z=2\pi\Omega_Z^{-1}$ within femtoseconds, accessible nowadays by femtosecond spectroscopy. 
This should be contrasted with the high-frequencies of roughly $10^{21}$ Hz, typical of the ZBW \cite{PhysRevD.23.2454} predicted by relativistic equations for electrons in vacuum, certainly not accessible by present experimental techniques. 
Thus, the quantum wave dynamics involves an interplay between the DIT and ZBW phenomena. 
Notice also that the ZBW  described by Eq.~(\ref{rho1_ZB}) exhibits a {\it quantum beat} effect governed by $\Omega_Z$, that we shall later discuss in our work.
\begin{figure}[H]
 \begin{center}
  {\includegraphics[angle=0,width=3.5in]{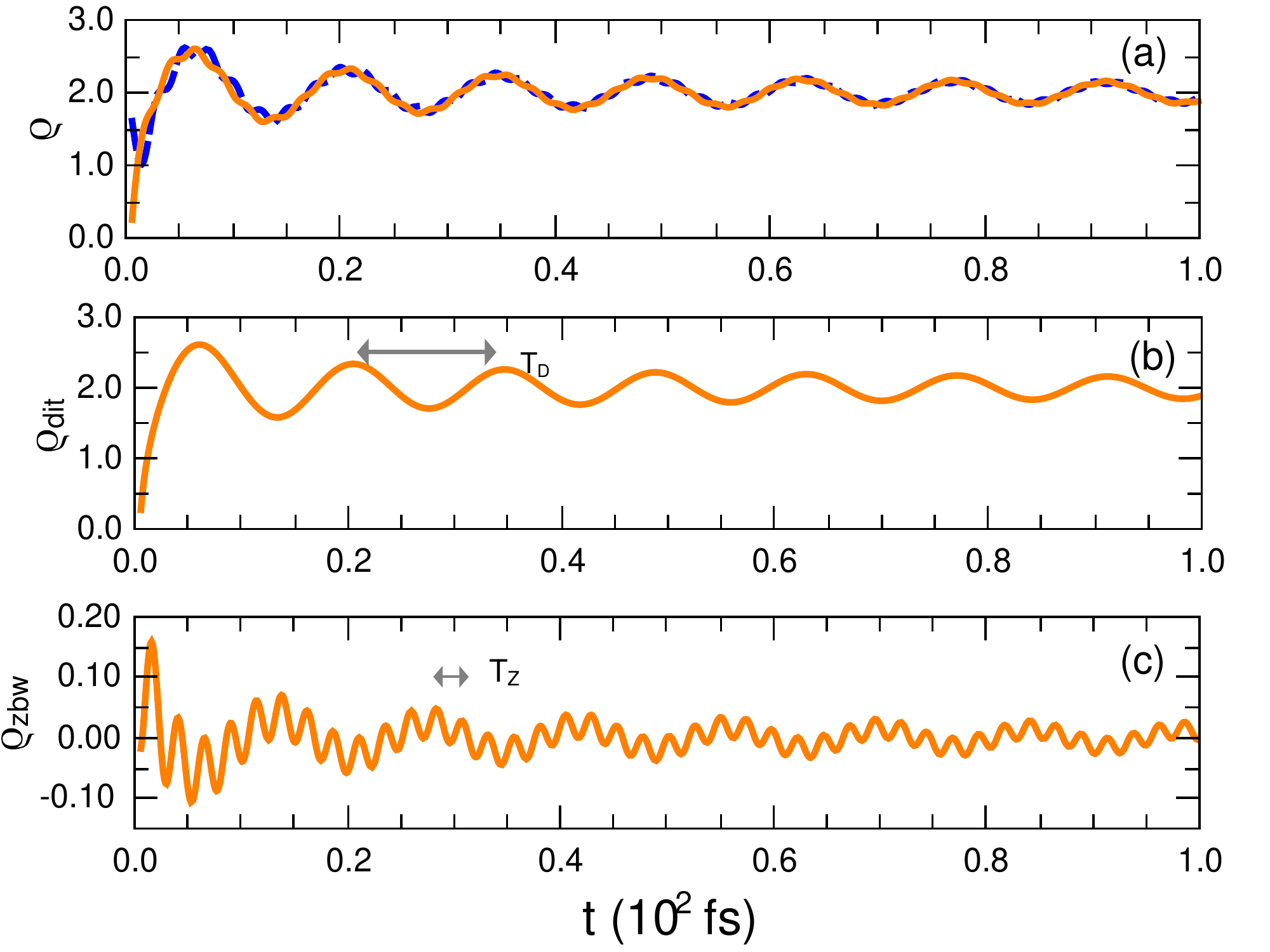}}
    \end{center}
     \caption{(a) The DIT and ZBW phenomena in the time-evolution of the exact
      $\rho$ (blue dashed line) using Eq.~(\ref{eq8b}), and the asymptotic  $\rho_a$ (orange solid line) using Eqs.~(\ref{rho0_DIT}) and (\ref{rho1_ZB}), for the case discussed in Fig.~\ref{fig1}(a). (b)  The DIT oscillations observed in (a) are described by $\rho_{dit}$ [Eq.~(\ref{rho0_DIT})]  with a period $T_D=141.2$ fs indicated in the plot, corresponding to a frequency $\Omega_D\sim 45.0$ THz. (c) The high-frequency oscillations  embedded in the DIT pattern of case (b)  are the manifestation of the ZBW effect, described by $\rho_{zbw}$ [Eq.~(\ref{rho1_ZB})], with a period $T_Z=29.25$ fs, and a frequency $\Omega_Z\sim 215.0$ THz.}
              \label{fig1_bis}
               \end{figure}
We stress that, although the DIT is a typical effect of  quantum shutter models for systems with non-zero mass, such as those described by Klein-Gordon \cite{mm52} or Dirac equations \cite{godoy16}, it also manifests itself in massless relativistic systems.
In our relativistic case, we argue that DIT and ZBW phenomena are a result of a momentum induced effective mass due to the transverse momentum component of the incident quantum wave, $k_y=k\sin\phi$. 
%
%
%
%
Therefore, We expect a strong dependence of the features of  quantum waves on the angle of incidence, $\phi$.
This is illustrated in Fig.~\ref{fig:fields}(a), where $\rho$ exhibits well defined DIT oscillations for incidences at $0<|\phi|< \pi/2$, as shown for example in Fig.~\ref{fig:fields}(c) and Fig.~\ref{fig:fields}(d).
Interestingly, the DIT effect is absent for the case with $\phi=0$ (normal incidence), as  shown in Figs.~\ref{fig:fields}(a), and \ref{fig:fields}(b).  
This peculiar result involves a ``massless" ($k_y=0$) system, and is consistent with the known fact that no DIT is observed  for the solution of the free quantum shutter problem for a one-dimensional wave equation \cite{mm52}.
We note that in the case discussed in Fig.~\ref{fig:fields}(d), it is difficult to resolve the ZBW and DIT frequencies. To understand this behavior, in Fig~\ref{fig:TyOmega}  we compare the periods and frequencies associated with these phenomena as  function of $\phi$. 
\begin{figure}[H]
 \begin{center}
  {\includegraphics[angle=0,width=2.6in]{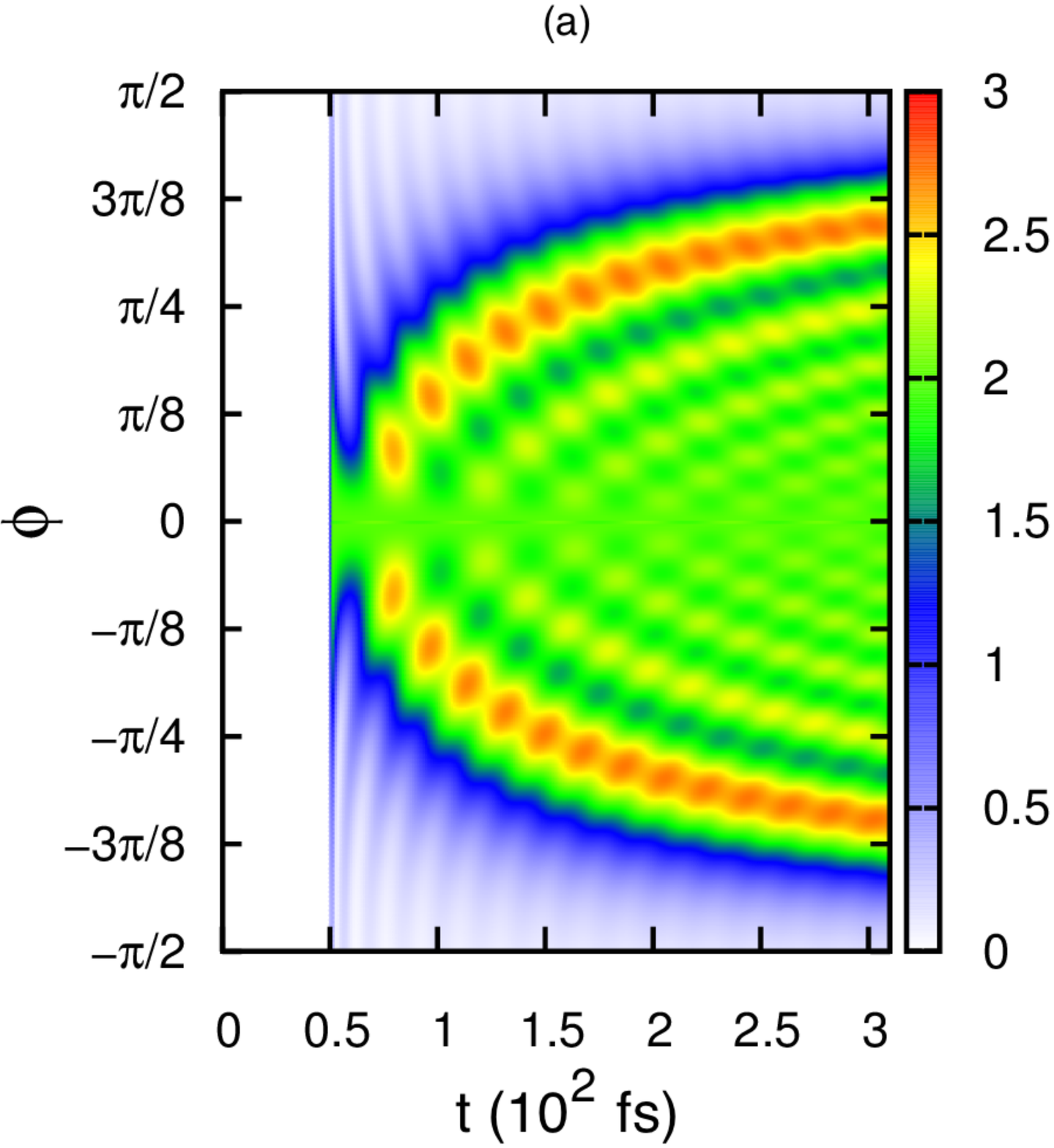}}
  {\includegraphics[angle=0,width=3.5in]{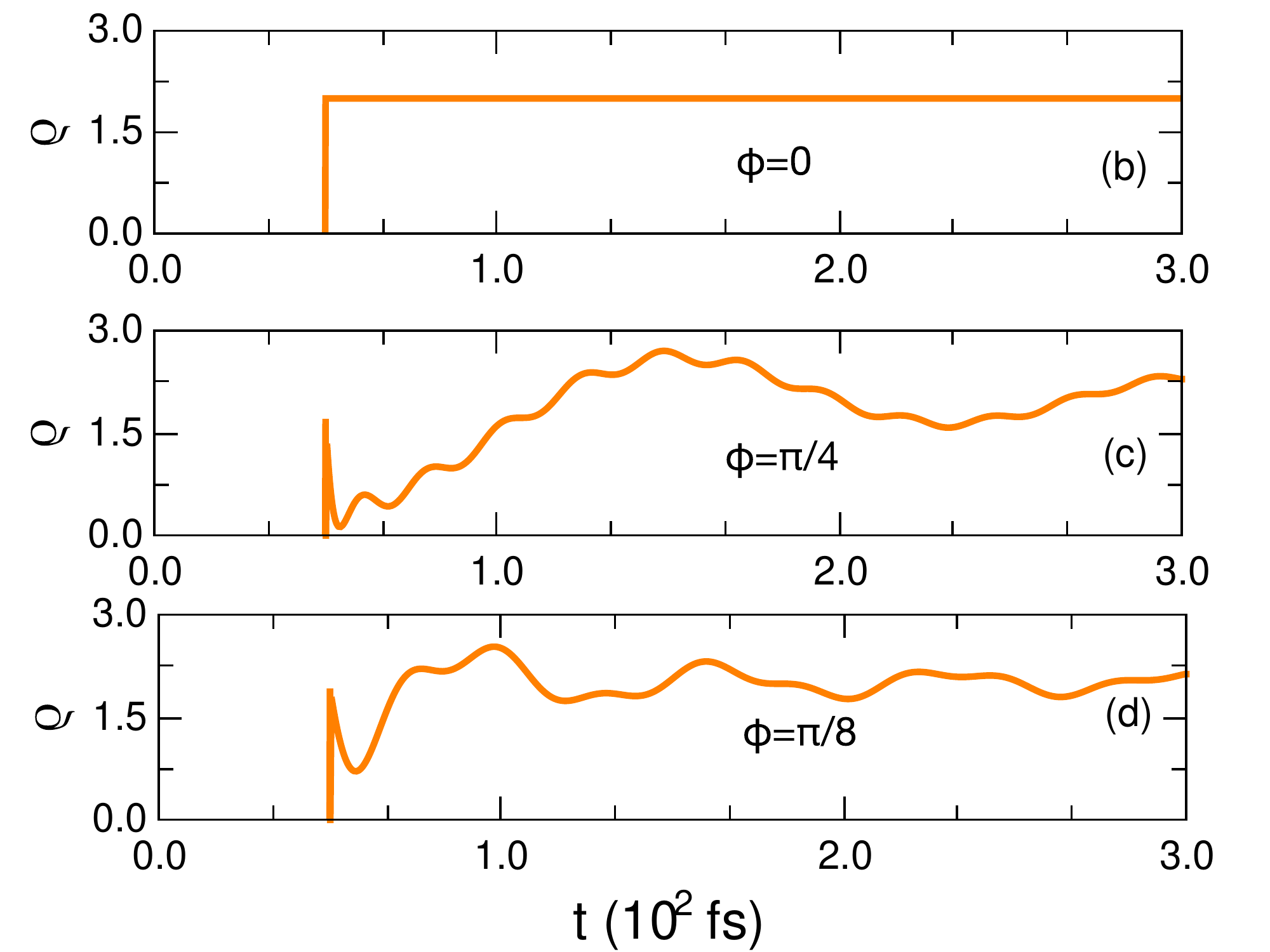}}
    \end{center}
     \caption {(a) Surface plot of  $\rho$ as  function of the incidence angle, $\phi$, and time, $t$. The lower panel shows the probability density as  function of the time $t$ for three different values of the incidence angle (b) $\phi=0$, (c) $\phi=\pi/4$, and  (d) $\phi=\pi/8$. Here we consider a fixed position $x=50.0$ nm, and the same parameters as in Fig.~\ref{fig1}.}
          \label{fig:fields}
           \end{figure}

From Fig.~\ref{fig:TyOmega} we can see that for $\phi\simeq\pi/8$, the DIT and ZBW frequencies are of the same order of magnitude {\it i.e.}  $\Omega_Z\simeq\Omega_D$, and thus difficult to resolve, as shown in Fig.~\ref{fig:fields}(d). 
Also, we note in Fig.~\ref{fig:TyOmega}(a) that for small values of $\phi$ we can easily resolve the frequencies since $\Omega_{Z}<\Omega_{D}$. We clarify that the DIT frequency, $\Omega_D$, obtained for the case $\phi=\pi/4$, also gives a good estimate of the time-oscillation frequencies for $\phi\ne\pi/4$.
As shown by our results, DIT is a robust effect observed for a wide range of values of $\phi$, and characterized by transient oscillations with frequencies that are currently accessible to experiments in the field of graphene.
\begin{figure}[H]
 \begin{center}
 {\includegraphics[angle=0,width=3.5in]{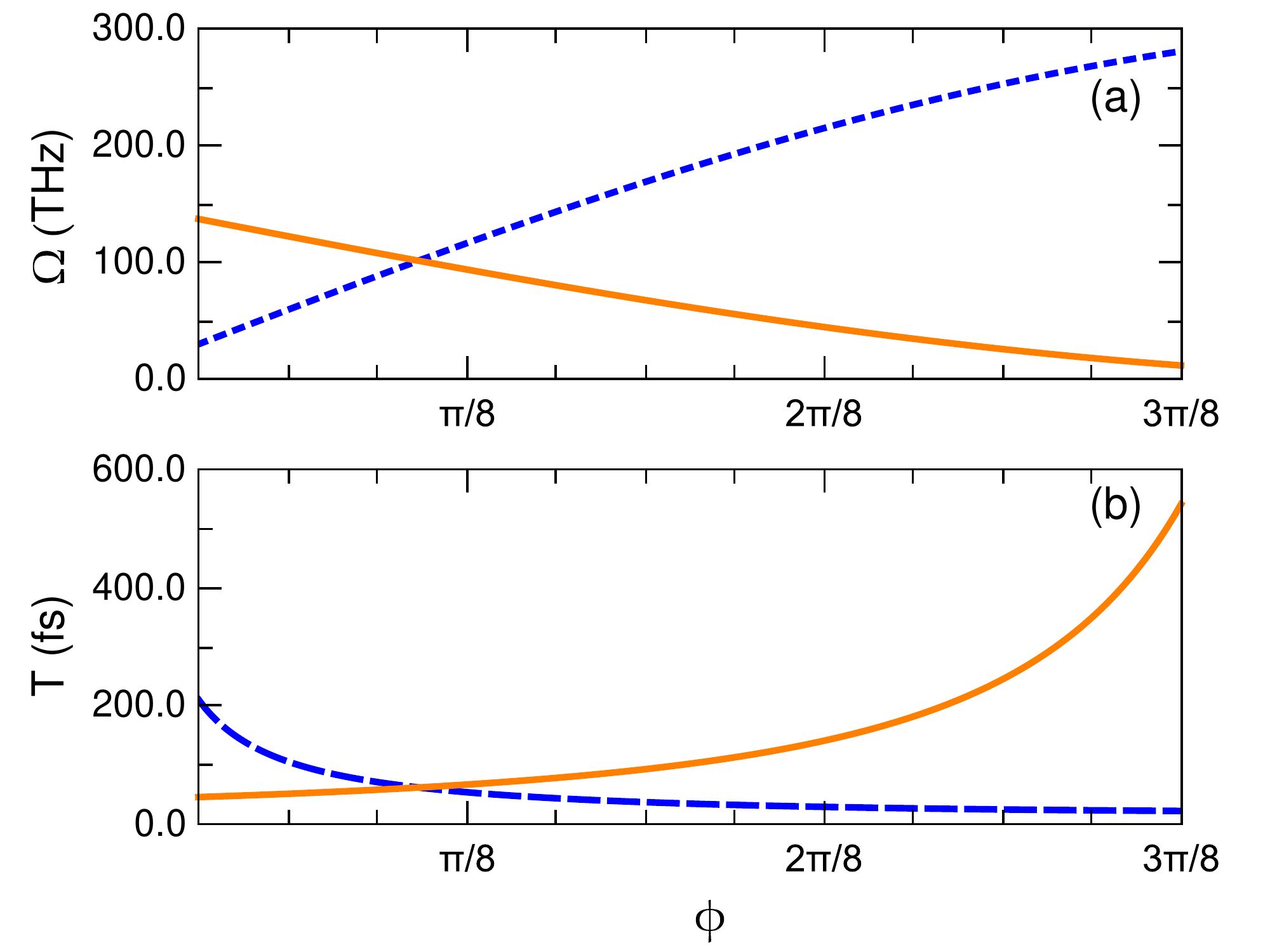}}
   \end{center}
    \caption {(a) Plots of the periods $T_D$ (orange solid line), and $T_Z$ (blue dashed line) as  function of $\phi$. Note that for this case  $\phi\simeq\pi/8$, $T_D\simeq T_Z$. (b) a) Plots of the frequencies  $\Omega_D$ (orange solid line), and $\Omega_Z$ (blue dashed line) as  function of $\phi$.  The frequency crossover  ($\Omega_Z=\Omega_D$) occurs at $\phi=\arctan(1/\sqrt{8})$. We use the parameters of Fig.~\ref{fig1}.}
        \label{fig:TyOmega}
         \end{figure}
\begin{figure}[H]
 \begin{center}
 {\includegraphics[angle=0,width=3.4in]{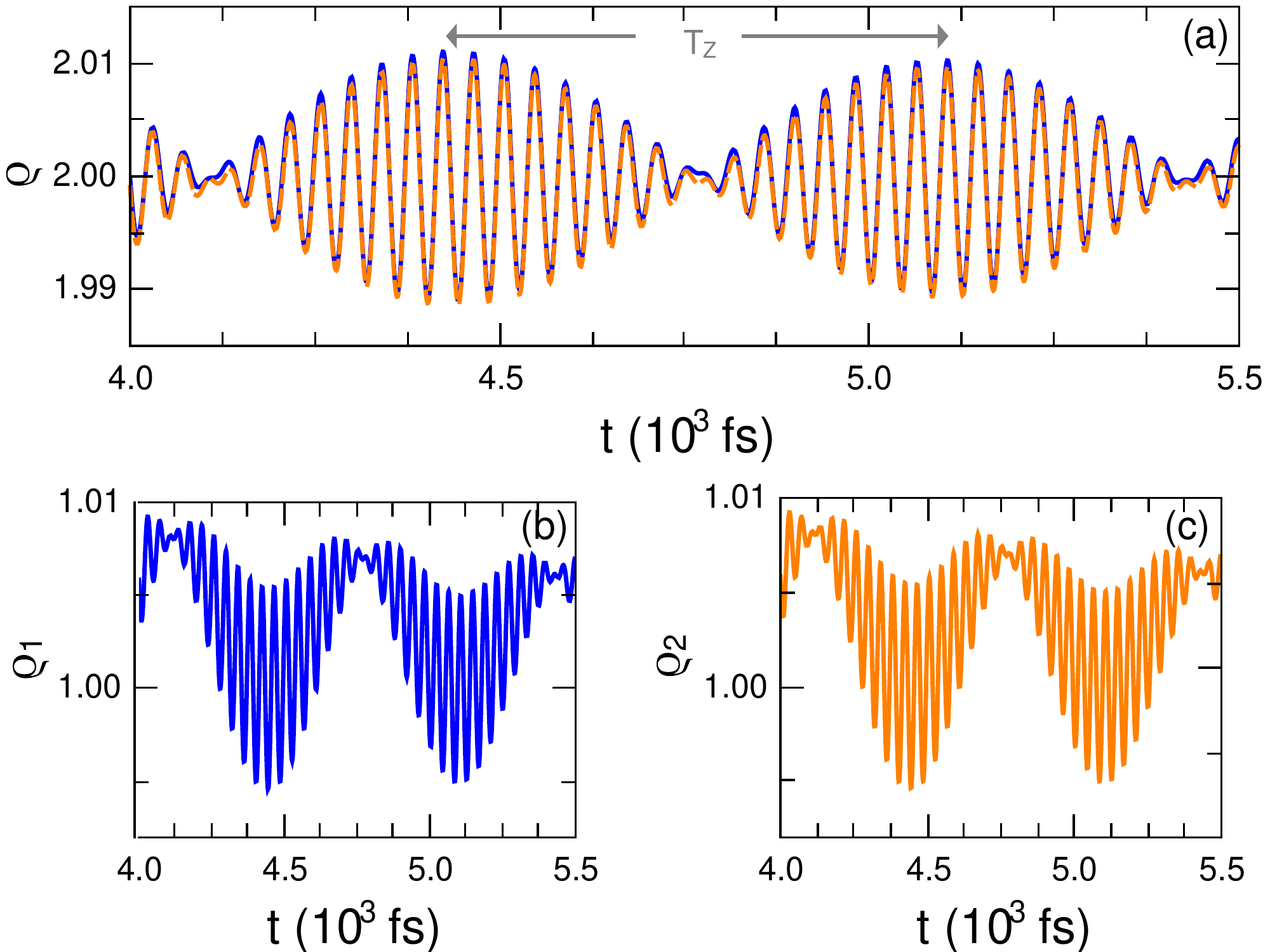}}
   \end{center}
    \caption{(a) {\it Quantum beat} phenomenon in the long-time behavior of $\rho$ using Eq.~(\ref{eq8b}) (blue solid line) at fixed position  $x=50.0$ nm, for the system discussed in Fig.~\ref{fig1}, for $\phi=\pi/100$. The frequency of the {\it quantum beats} is the governed by the ZBW frequency, $\Omega_Z=9.5$ THz, with a period $T_Z=658.4$ fs, indicated in the plot. The asymptotic $\rho_s$ (orange dashed line) is included for comparison, and almost completely overlaps with the exact solution. We also show the high-frequency beatings induced by the two asymptotic contributions (b) $\rho_1$ (solid blue line)  and (c) $\rho_2$ (orange solid line). }
            \label{fig1_ZB_pi100}
             \end{figure}

\textit{Zitterbewegung effect.} 
Before exploring the subject, we emphasize that 
ZBW is usually studied by means of the time-evolution of the expectation value of the electrons position for wave packets within  Heisenberg's picture. 
Moreover, it  has been recognized that the ZBW effect is of a transient nature \cite{PhysRevB.76.195439}.
In this regard, we are proposing  an alternative time-dependent approach to address the issue of ZBW, by exploring the transient behavior of the probability density at long-times ($t\gg t_F$).
So far we have studied the interplay between DIT and ZBW phenomena, which can appear at different frequencies. 
We have also discussed in Fig.~\ref{fig:TyOmega} that there is a regime of low-incidence angles,  where the ZBW may be the dominant effect since $\Omega_{Z}<\Omega_{D}$. 
In Fig.~\ref{fig1_ZB_pi100} we consider the time-evolution of the probability density for the system discussed in Fig.~\ref{fig1}, for small values of $\phi$ {\it i.e.} near the normal incidence condition ($\phi=0$). 
Notice that $\rho$ exhibits a series of {\it quantum beats} that modulate  high-frequency oscillations,  similar to the superposition phenomenon of quantum waves with different frequencies.
To identify the  underlying  superposition that originates the {\it quantum beats}, we derive an asymptotic formula for the long-time behavior of $\rho$, for small values of $\phi$.
Therefore, by analyzing the asymptotic behavior of the solutions $\phi_{\pm}$ [Eq.~(\ref{simplifbis2})] with the help, as before, of the asymptotic formula for the Bessel function $J_n(z)$ for large values of the argument $z$, we obtain
$\rho_s=\rho_{1}+ \rho_{2}$, with $\rho_{1}=\rho^{+}$, and $\rho_{2}=\cos\phi\,\rho^{-}$, with
\begin{eqnarray}
\rho^{\pm}&=& 1+\sqrt{2} z_+^{-1}\gamma t^{-1/2}\sin \left(k_x x-\omega t\right)\sin\left[\left(\Omega_Zt-\pi/2\right)/2\right] \nonumber \\ \label{rhocos}
&\pm& \left(\gamma^2/16 \,t \right)\left(1+\sin \Omega_Z t\right).
\end{eqnarray}
%
From Eq.~(\ref{rhocos}), the ZBW  emerges in the probability density as a series of {\it quantum beats} of frequency $\Omega_Z$,  due to a superposition of  sinusoidal quantum waves, with an amplitude that decays as $t^{-1/2}$.
The asymptotic  $\rho_s$ fits our exact result, as shown in Fig.~\ref{fig1_ZB_pi100}(a). Moreover, the beating frequency can be obtained by inspection of the ZBW period $T_Z$ in Fig.~\ref{fig1_ZB_pi100}(a) of the order of hundreds of femtoseconds, which yields frequencies in the range of tens of terahertz. 
The interpretation of these oscillations has been overlooked in studies addressing the problem of transients in Dirac theory \cite{moshrmf52,godoy16}.
In our case, it is striking that an effect typical of matter-waves occurs in a system where the dynamics is associated to massless fermions.

\begin{figure}[H]
 \begin{center}
 {\includegraphics[angle=0,width=3.4in]{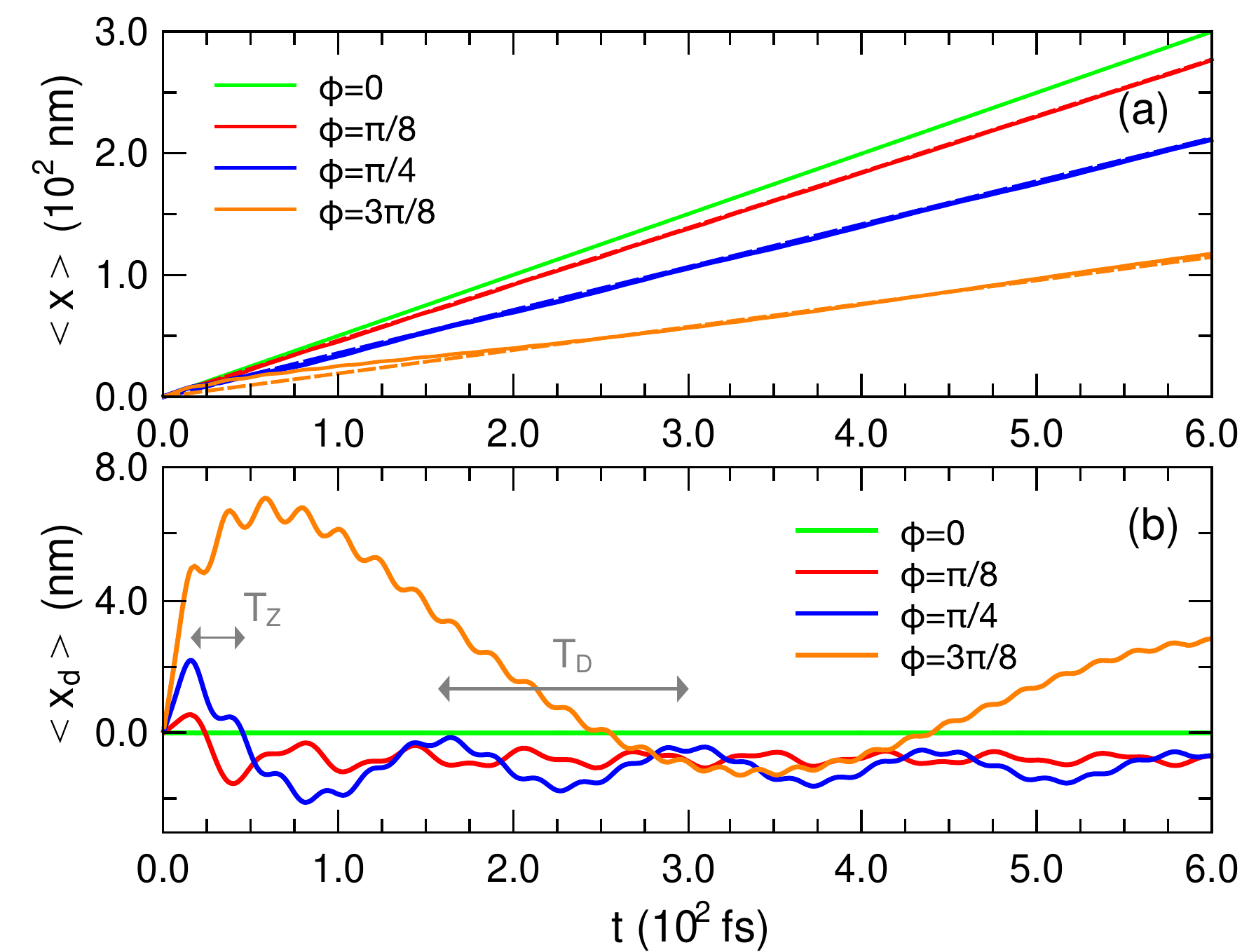}}
   \end{center}
    \caption{ (a) Time-dependence of $\langle x\rangle$ [Eq.~(\ref{eqa6.1})] (solid lines), and  $\widetilde{x}$ [Eq.~(\ref{eqa6.3})] (dashed lines), for different values of the incidence angle, $\phi$.  The expectation value $\langle x\rangle$ slightly oscillates around the $\widetilde{x}$, which may indicate the presence of ZBW and DIT phenomena, for different values of $\phi$. (b) The average $\langle x_d \rangle$ shows an interplay between ZBW and DIT. We include in the plot the DIT and ZBW periods, $T_D=141.2$ fs, and $T_Z=29.5$ fs, respectively, for the particular case of $\phi=\pi/4$.}
        \label{zitter_1}
         \end{figure}

To further explore the features of the ZBW effect in different time-regimes, we also analyze the transient behavior of the expectation value of the position operator, $\langle x\rangle$, using the exact  $\rho$ calculated from Eq.~(\ref{eq8b}).  
Therefore, we define $\langle x\rangle$ for $x>0$ as,
\begin{equation}
\langle x\rangle=\left(\int_{0}^{v_F t}x\rho(x,t;\phi)dx\right)\Bigg/\left(\int_{0}^{v_F t}\rho(x,t;\phi)dx\right),
\label{eqa6.1}
\end{equation}
where we emphasize the explicit $\phi$ dependence of $\rho$. The integration limits are chosen to fulfill Einstein's causality {\it i.e.} %
$\rho$ is non-zero for $x<v_Ft$.
We also introduce, for comparison purposes, the expectation value of position, $\widetilde{x}$, associated to the free-propagation of massless electrons, represented by a plane wave that propagates with velocity $v_F$ in the $\hat{k}=\bm{k}/k$ direction, where $\bm{k}$ is the wave vector.
Thus, $\widetilde{x}$ for $x>0$ can be expressed as,
\begin{eqnarray}\label{eqa6.2}
\widetilde{x}=\left(\int_{0}^{v_xt}x\rho~dx\right)\Bigg/ \left(\int_{0}^{v_xt}\rho~dx\right).
\end{eqnarray}
Since $\rho$ is a constant over all the region $0<x<v_Ft$, and the wavefront propagation speed is $v_x=v_F \cos\phi$, by integrating Eq.~(\ref{eqa6.2}), we obtain, 
\begin{eqnarray}\label{eqa6.3}
\widetilde{x}=v_F t \cos(\phi)/2. 
\end{eqnarray}
%
In  Fig.~\ref{zitter_1}(a) we compare the  time-dependence of $\langle x\rangle$ [ Eq.~(\ref{eqa6.1})], and  $\widetilde{x}$ [Eq.~(\ref{eqa6.3})], for different values of $\phi$.  
We show in Fig.~\ref{zitter_1}(a) that for the case $\phi=0$ ($k_y=0$), the time-dependence of $\widetilde{x}$ (red dashed line),  and $\langle x\rangle$ (red solid line) is characterized by a straight line, in agreement with the result reported in Ref. \onlinecite{Gerri} for massless particles. 
We can also see in Fig.~\ref{zitter_1}(a) that for values of the incidence angle in the range $0<\phi<\pi/2$, the average position $\langle x\rangle$ (solid line) displays weak oscillations around the corresponding free-type expectation value,  $\widetilde{x}$ (dashed line). 
We argue that  $\widetilde{x}$ hinders the oscillatory behavior barely exhibited by $\langle x\rangle$, so in Fig.~\ref{zitter_1}(b) we proceed to remove from $\langle x\rangle$ the contribution of $\widetilde{x}$, by  computing the difference, $\langle x_d\rangle=(\langle x\rangle -\widetilde{x})$.   
As a result, $\langle x_d\rangle$ features a complex oscillatory pattern, reflecting the same interplay between the DIT and ZBW phenomena, that we found by exploring the transient behavior of $\rho$. 
For example, in the case $\phi=\pi/4$ shown Fig.~\ref{zitter_1}(b), $\langle x_d\rangle$ exhibits the same DIT oscillations with a period $T_D$, as well as the embedded secondary high-frequency ZBW oscillations with period $T_Z$, 
as discussed in Fig.~\ref{fig1_bis}. 
%

\textit{Conclusions.} 
We explore the transient dynamics of cut-off plane waves for two-dimensional massless Dirac-fermions, by deriving an exact analytical solution to a Dirac-type equation within a quantum shutter setup. 
Our time-dependent approach is developed for a graphene \cite{RevModPhys.81.109,Novoselof05}  monolayer in the low-energy regime, and can be readily applied to other promising systems such as 8-$Pmmn$ borophene \cite{lopezbonilla16,peng16,Zabolotskiy16,verma17}. 
We find that the probability density is characterized by an interplay of two distinctive oscillatory phenomena in the time-domain: 
(i) a {\it time-diffraction} effect of frequency $\Omega_D$, 
similar to the DIT \cite{mm52} phenomenon for Schr\"odinger free matter-waves; 
(ii) a ZBW effect which manifest itself as transient oscillations of frequency $\Omega_Z$, embedded in the DIT-like profile. 
We show that these transient effects arise due to the non-zero transverse momentum of the initial quantum wave, $k_y$, that acts as an effective mass of the system. 
This acquired property of mass can be manipulated by simply tuning the incidence angle, $\phi$, which allows us to adjust the frequencies  $\Omega_D$, and $\Omega_Z$, and control to what extent the DIT or ZBW phenomena dominate the dynamics.
In particular, it is shown that near a normal incidence condition, there is a time-regime where the ZBW emerges as a series of {\it quantum beats} in the probability density, with a beating frequency, $\Omega_Z$,  and an amplitude that decays as $t^{-1/2}$.
We have found system configurations where the frequency of these transient effects are in the  terahertz regime, accessible to nowadays experimentalists by using femtosecond spectroscopy. 
Our results may be useful in experimental setups involving the study of transient oscillations in quantum simulations like those of Ref. \onlinecite{Gerri}, where the terahertz regime is accessible to investigate the DIT and ZBW phenomena.

We also hope that our results may stimulate further studies on transient phenomena for cut-off quantum waves in two-dimensional materials, particularly in systems that involve different potential profiles or initial conditions, as for example by tailoring initial states as a superposition of plane waves to construct finite width wave packets.

\section*{Acknowledgements}
We acknowledge useful discussions with David Ruiz-Tijerina and Mahmoud Asmar. R.C. acknowledges useful discussions with V.G. Ibarra-Sierra and J.C. Sandoval-Santana. E.C. was fully supported by Beca-PRODEP. The authors also acknowledge financial support from FC-UABC under Grant PROFOCIE 2018.

\bibliographystyle{apsrev4-1}
\bibliography{biblio.bib}

\end{document}